\documentclass[prl,twocolumn,superscriptaddress,nofootinbib,floatfix]{revtex4-1}
\usepackage{epsfig}
\usepackage{graphicx}
\usepackage{palatino}
\usepackage[english]{babel}
\usepackage{hyphenat}
\usepackage{amsmath}
\usepackage{amssymb}
\usepackage{mathtools}
\usepackage{mathrsfs}
\usepackage{slashed}
\usepackage{epstopdf}
\usepackage{xcolor}
\usepackage{braket}
\usepackage{booktabs}
\definecolor{lcolor}{rgb}{0.,0.0,0.}
\definecolor{citcolor}{rgb}{0,0.,0.5}
\usepackage[breaklinks,colorlinks,urlcolor=blue,citecolor=blue,linkcolor=blue]{hyperref}
\usepackage{multirow}
\usepackage{ltablex}
\usepackage{soul}

\newcommand{\beq}{\begin{eqnarray}}
\newcommand{\eeq}{\end{eqnarray}}

\newcommand{\bem}{\begin{multline}}
\newcommand{\eem}{\end{multline}}
\newcommand{\beg}{\begin{gather}}
\newcommand{\eeg}{\end{gather}}

\newcommand{\nn}{\nonumber\\}

\newcommand{\ben}{\begin{eqnarray*}}
\newcommand{\een}{\end{eqnarray*}}

\def\cP{{\cal P}}

\newcommand{\secn}[1]{Section~1}
\newcommand{\appn}[1]{Appendix~1}

\long\def\comment#1{ }

\def\and{\quad\text{and}\quad}

\def\0{{\boldsymbol 0}}

\def\0{{\boldsymbol 0}}

\begin{document}

\title{Leading order track functions in a hot and dense QGP}

\author{João Barata}
\email[]{jlourenco@bnl.gov}
\affiliation{Physics Department, Brookhaven National Laboratory, Upton, NY 11973, USA}
\author{Robert Szafron}
\email[]{rszafron@bnl.gov}
\affiliation{Physics Department, Brookhaven National Laboratory, Upton, NY 11973, USA}

\begin{abstract}
We study the modifications to the fragmentation pattern of partons into charged particles in the presence of a hot and dense Quark Gluon Plasma. To this end, we analyze the perturbative renormalization group equations of the track functions, which describe the energy fraction carried by charged hadrons. Focusing on pure Yang-Mills theory, we compute the lowest order moments of the medium-modified track functions, which are found to be sensitive to the reduced phase space for emissions in the medium and to energy loss. We use the extracted moments to calculate the Energy Energy Correlator (EEC) on tracks in the collinear limit. The EEC on medium-evolved tracks does not differ qualitatively from the EEC on vacuum tracks despite being 
sensitive to the color decoherence transition and suppressing the distribution due to quenching, as seen in other jet observables.
\end{abstract}

\maketitle

\textit{Introduction:}
Jets are an ideal tool to study the time evolution and spatial structure of the Quark Gluon Plasma (QGP) produced in heavy ion collisions. When using jets in this manner, two problems need to be addressed: how the QCD matter modifies the jets' structure~\cite{Mehtar-Tani:2013pia,Qin:2015srf} and which observables best reveal the medium imprints in the measured distributions ~\cite{Cunqueiro:2021wls,Apolinario:2022vzg}.

 Regarding the latter point, we address an open question about Energy Energy Correlator (EEC) measurements inside jets, i.e. in the collinear limit, traveling through the QGP~\cite{Andres:2023xwr,Andres:2022ovj,Barata:2023zqg,Barata:2023vnl,Barata:last_paper_2023,Yang:2023dwc}. The EEC is the simplest correlation function constructed from quantum field theory light-ray operators measuring the asymptotic energy flow passing through idealized calorimeter cells, see e.g.~\cite{Chen:2023zzh, Chen:2020vvp, Hofman:2008ar}.
It can be written in terms of the inclusive two-particle cross-section as~\cite{PhysRevD.17.2298}
\begin{align}
\label{eq:eec-definition}
\frac{d\Sigma^{(n)}}{d \chi} &= \sum_{\lbrace i, j\rbrace} \int d\sigma \, \frac{E_i^n E_j^n}{p_t^{2n}} \delta(\vec{n}_i\cdot \vec{n}_j-\cos\chi) \, ,
\end{align}
where $\vec n_i$ denotes the spatial direction of the measured energy flow, $E_i$ is the energy of particle $i$, $p_t$ is the total energy of the system, and the sum is taken over all particles in the final state. The integer $n\geq 1$ is introduced as an effective way to further suppress contributions to the EEC from soft emissions.

Values of $n>1$ are especially relevant in the heavy ion context where uncorrelated soft sources contaminate jets. In such cases, the EEC is infrared-and-collinear (IRC) unsafe, 
and Eq.~\eqref{eq:eec-definition} cannot be evaluated perturbatively. Nonetheless, one can separate the perturbative and non-perturbative contributions to the EEC~\cite{Chen:2020vvp,Li:2021zcf}. The matching coefficients between the parton level EEC and the full distribution are given by the moments
of the so-called track functions (TFs)~\cite{Chang:2013rca}.\footnote{This issue can also be avoided by defining the EEC on objects which are naturally IRC safe~\cite{Barata:last_paper_2023}.}

\begin{figure}[h!]
    \centering
    \includegraphics[width=.4\textwidth]{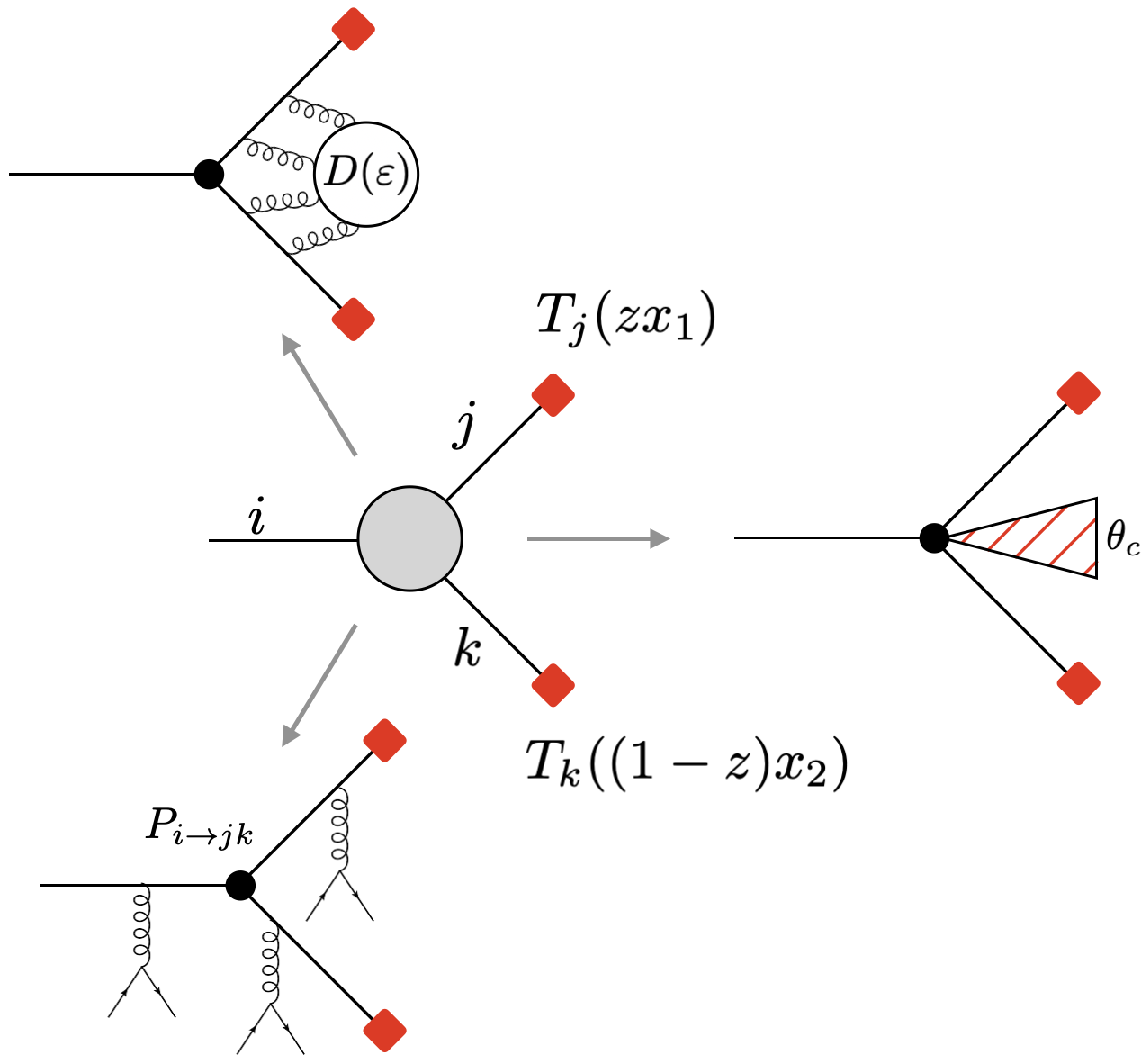}
    \caption{Diagrammatic  representation of the LO evolution kernel (see Eq.~\eqref{eq:track_LO_vac}) and the different medium modifications we take into account: energy loss (top left), induced radiation (bottom left), and reduced phase space (right).}
    \label{fig:diagram}
\end{figure}

Several theoretical and phenomenological aspects of the track functions have been explored for pure vacuum evolution, see e.g.~\cite{Chang:2013rca,Chen:2020vvp,Li:2021zcf, Chang:2013iba,Lee:2023npz}. They have not been studied in the heavy ion context, and the in-medium EEC calculations have so far neglected the matching to the hadronic level observable~\cite{Andres:2023xwr,Barata:last_paper_2023}. The goal of this work is to provide the first phenomenologically driven study of the medium effects on the track functions and explore how these affect the jet EEC on tracks.

\textit{Leading order vacuum track functions:} Tracks functions were first introduced a decade ago~\cite{Chang:2013rca, Chang:2013iba}, allowing to compute jet observables based on hadrons with specific quantum numbers in a theoretically well-defined way; in particular TFs opened the possibility to compare theory calculations to experimental measurements on charged tracks. In the simplest case, a track function $T_f(x)$ 
describes the fragmentation of a parton of flavor $f$ into a subset of hadrons\footnote{Here, we have in mind the fragmentation into charged hadrons, but the calculation is valid for other conserved quantum numbers.} carrying an energy fraction $x$. Similarly to collinear fragmentation functions (FFs)~\cite{Collins:1981uw}, TFs are intrinsically non-perturbative. However, unlike FFs, they obey a highly non-linear renormalization group (RG) evolution~\cite{Jaarsma:2022kdd,Li:2021zcf,Chen:2022muj}, more closely related to multi-hadron fragmentation~\cite{Sukhatme:1980vs}.\footnote{A detailed description of the RG group for track functions has recently become available beyond leading order (LO)~\cite{Li:2021zcf}.} At leading order (LO) in the strong coupling $\alpha_s$, $T_i(x)$ satisfies the RG evolution equation~\cite{Chang:2013rca,Jaarsma:2022kdd}
\begin{align}\label{eq:track_LO_vac}
 &\mu\frac{d T_i(x,\mu)}{d \mu}  = \sum_{j,k}\int_0^1 dz\, \frac{\alpha_s}{2\pi} P_{i\to jk}(z) \nn 
 &\times \int_{x_1,x_2} T_j(x_1,\mu) T_k(x_2,\mu) \delta(x-zx_1-(1-z)x_2)\, ,
\end{align}
where $P_{i\to jk}$ are the LO regularized QCD splitting functions \cite{Altarelli:1977zs}, and the evolution is expressed in terms of the renormalization scale $\mu$, see Fig.~\ref{fig:diagram}. The DGLAP evolution~\cite{Dokshitzer:1977sg,Altarelli:1977zs,Gribov:1972ri} can be recovered by restricting Eq.~\eqref{eq:track_LO_vac}
to a one-body evolution, i.e., when it is fully inclusive by replacing $T_k(x) \to \delta(x)$~\cite{Chen:2022pdu}, for a subset of particles.\footnote{ We use the notation $\int_x$ to denote the integration over the full phase space in the energy fraction $x$.}

The RG equation simplifies for the track functions' moments\footnote{The dependence on the evolution scale $\mu$ is made implicit.}
\begin{align}\label{eq:moment_def}
   T_i^{[N]} \equiv \int_0^1dz\, z^N T_i(z)\, .
\end{align}
In Yang-Mills (YM) theory, it follows from Eq.~\eqref{eq:track_LO_vac} that the moments satisfy 
\begin{align}\label{eq:track_general}
  &\mu\frac{d T^{[N]}}{d \mu}  =  \frac{\alpha_s}{ 2\pi}\int_0^1 dz P(z)   \nn 
  &\times \int_{x_1,x_2} T(x_1,\mu) T(x_2,\mu)  (zx_1+(1-z)x_2)^N \, , 
\end{align}
where $P(z)\equiv P_{g\to gg}(z)= 2N_c (z/(1-z)_++(1-z)/z+z(1-z))+11N_c/6\,  \delta(1-z)$ and $T(z)=T_g(z)$. Using conservation of energy to impose the sum rule $T^{[0]}=1$, and the symmetry of the splitting function under the integral after shifting the poles to $z=1$, i.e., $\int_z P(z)(1-z) = \int_z P(z)z$,  the leading moments' evolution reads
\begin{align}\label{eq:T1_moment_vac_gamma}
  &\mu\frac{d T^{[1]}}{d \mu}  =-\frac{\alpha_s}{\pi} \gamma(2)  T^{[1]}\, ,\nn 
    &\mu\frac{d T^{[2]}}{d \mu} = -\frac{\alpha_s}{\pi}\left( \gamma(3)  T^{[2]} -(\gamma(2)-\gamma(3)) T^{[1]}T^{[1]} \right) \, .
\end{align}
Here, we have introduced the anomalous dimensions
\begin{align}\label{eq:AD}
  \gamma(j) &\equiv  -\int_0^1 dz\, z^{j-1} P(z) \, , 
\end{align}
 such that in the vacuum $\gamma_{\rm vac}(j)=- 11N_c/6-2 N_c (1/(j^2-j)+1/((j+1)(j+2)) - \psi(j+1)-\gamma_E] $, with $\psi(j) \equiv \Gamma'(j)/\Gamma(j)$ the digamma function and $\gamma_E\approx0.577$ the Euler-Mascheroni constant. Equations~\eqref{eq:T1_moment_vac_gamma} can be further simplified in YM theory, where $\gamma(2)=0$. This sum rule follows from energy conservation; as a result, it provides a non-perturbative constraint that must also be satisfied for in-medium evolution, as long as energy loss effects are not present. The related RG invariance under the shift symmetry $T(x)\to T(x+b)$~\cite{Li:2021zcf,Jaarsma:2022kdd} constrains higher moments' evolution, but there are no further non-trivial constraints in YM theory.

\textit{Medium modifications to the track functions:} We assume that the partonic cascade occurs in a homogeneous and isotropic gluonic QCD medium, with a sizeable longitudinal extension $L$.\footnote{For recent developments in the description of fragmentation in more realistic matter models see e.g.~\cite{Sadofyev:2021ohn,Kuzmin:2023hko,Barata:2023qds}.
} We take the interactions with the underlying matter to be dominated by multiple soft gluon exchanges, see  Fig.~\ref{fig:diagram}, which are captured in the BDMPS-Z/ASW phenomenological model~\cite{Zakharov:1996fv,Wiedemann:1999fq,Baier:1994bd,Salgado:2003gb}. Large momentum exchanges with the medium are neglected; the discussion can be easily extended to such cases.

We first consider the hardest/earliest branchings in the cascade, which are primarily vacuum-like~\cite{Caucal:2018dla}.  The phase space for these emissions, at leading logarithmic accuracy, can be captured by introducing the semi-classical formation time $t_f=2/(z(1-z)\theta^2 p_t)$; the radiation pattern is modified depending on the ordering of $t_f$ with respect to the characteristic time scale generated by the medium, $t_c$. In particular, the vacuum-like region corresponds to $t_f<t_c$~\cite{Caucal:2018dla}. 
Alternatively, we can use angles to define the vacuum-like sector as $\theta>\theta_c$, where $\theta_c$ is the characteristic medium angular scale.
For the matter model considered, the characteristic angular scale is $\theta_c\equiv 2/\sqrt{\hat q_{\rm eff} L^3 }$~\cite{Casalderrey-Solana:2012evi,Mehtar-Tani:2012mfa}, with $\hat q_{\rm eff} = \hat q (1-z+z^2)\equiv \hat q f(z)  $ the effective jet quenching transport coefficient in YM theory, and $\hat q$ is the scalar adjoint transport coefficient; the corresponding characteristic time scale is $t_c= \sqrt{2z(1-z)p_t/\hat q_{\rm eff}}$~\cite{Mehtar-Tani:2017web,Caucal:2018dla}. 

We consider two different phase spaces for vacuum emission, differing mainly in the inclusion of out-of-the-medium fragmentation. Since the cascade still obeys vacuum evolution, the functional form of Eq.~\eqref{eq:track_general} is not affected, and the phase space modification can be formally absorbed into the anomalous dimension~\cite{Barata:last_paper_2023} 
  \begin{align} \label{eq:lllooo}
    & \gamma(j,\mu) = - \int_0^1 dz \, P(z)\Theta_{\rm PS}(\mu, z) z^{j-1} \, ,
  \end{align}
where $\Theta_{\rm PS}$ denotes the allowed phase space for fragmentation. The first model only allows for branchings to occur if they are sufficiently short-lived and wide-angled to be resolved~\cite{Mehtar-Tani:2017web}: $\Theta^{(1)}_{\rm PS} = \Theta(\mu-p_t \theta_c) \Theta(t_c-t_f)$,
where the scale $\mu=p_t \theta$. The second phase space we study allows fragmentation in the region $t_f>L$~\cite{Caucal:2018dla}: $ \Theta^{(2)}_{\rm PS}=1-\Theta(t_f-t_c) \Theta(L-t_f)$. Both phase spaces are formally only well described in the $z\ll 1$ limit, but we naively extend them to full kinematics here. A study of the phase space structure at finite energy fractions is still missing from the literature.

\begin{figure}[h!]
    \centering
    \includegraphics[width=.45\textwidth]{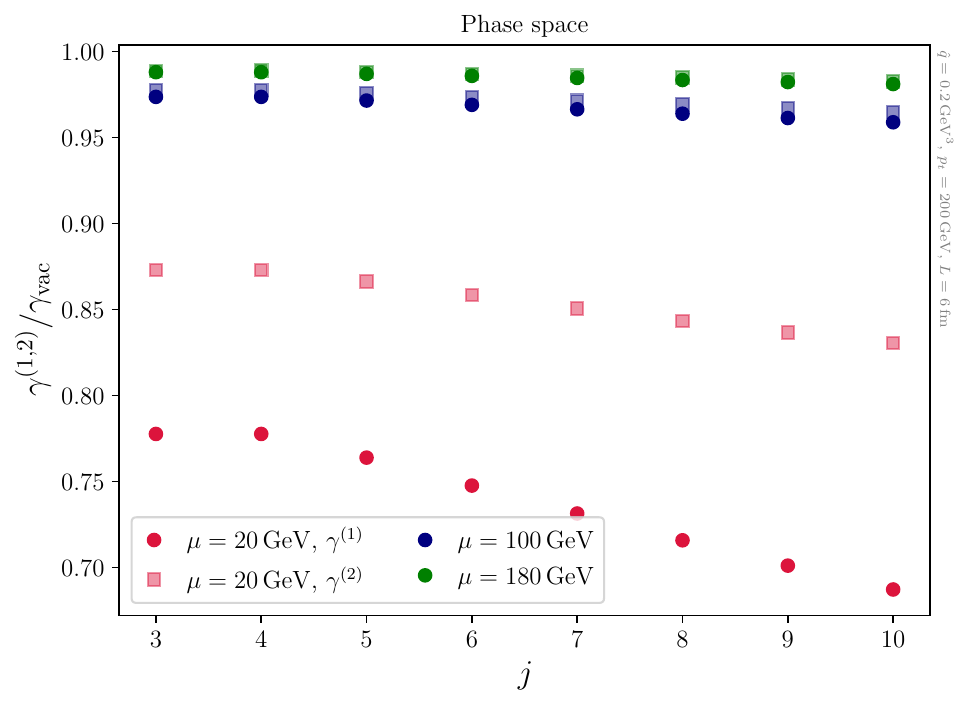}
    \includegraphics[width=.45\textwidth]{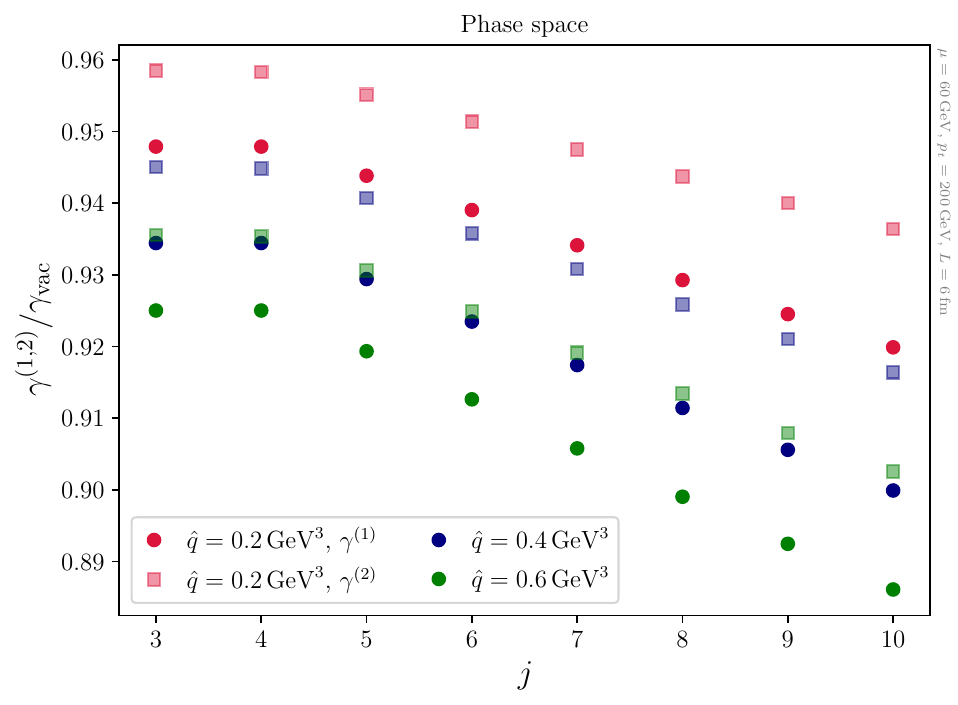}
    \caption{Evolution of the anomalous dimensions as a function of the evolution scale $\mu$ and the jet quenching parameter $\hat q$.}
    \label{fig:gamma_PS}
\end{figure}

To better understand the medium modifications, we first compute the anomalous dimensions in the soft limit. Using the soft (BFKL) form of the vacuum anomalous dimension, $\gamma^{\rm BFKL}_{\rm vac}(j)= \frac{2N_c}{1-j}$, we find that:
\begin{align}\label{eq:lloppp}
    \gamma^{(i)}(j,\mu) &=\gamma^{\rm BFKL}_{\rm vac}(j)\left(\lambda_{i+}^{j-1}-\lambda_{i-}^{j-1} + \delta_{i,2}\right)\, , 
\end{align}
where $i=1,2$ denotes the phase space model, $\lambda_{1-} = \lambda_{2+} = \sqrt[3]{2\hat q p_t/\mu^4 } $, $\lambda_{1+} =1$ , $\lambda_{2-} =2p_t/(\mu^2 L)$, and the phase space constraints depending only on $\mu$ are left implicit. Inside the medium, the BFKL pole vanishes since soft gluons cannot go on-shell. Also, Eq.~\eqref{eq:lloppp} ensures the strongest medium modifications occur at smaller integer values $j>1$.  In Fig.~\ref{fig:gamma_PS}, we show the anomalous dimensions for both models at finite $z$, while imposing the sum rule $\gamma(2)=0$. On top, we show the evolution with the scale $\mu$. As expected, if $\mu\approx p_t$ (i.e. $\theta\approx 1$), we recover the vacuum result as there is no loss of phase space, i.e., highly energetic jets are not sensitive to the medium. Taking the evolution scale down, the medium reduces the phase space available to radiate, and thus, the anomalous dimension also decreases. The significant difference between the two models resides in the $\mu\to 0$ region (at fixed $p_t$): while $\gamma^{(1)}\to 0$,  $\gamma^{(2)}\to \gamma_{\rm vac}$ since at small angles out-of-the-medium emissions are included, see Fig.~3 in~\cite{Barata:last_paper_2023}. At fixed $\mu$, the denser the medium, the larger the modification, as expected. These are more prominent for $\gamma^{(1)}$, where no out-of-the-medium evolution is incorporated, and for larger $j$. The latter originate from intermediate values of $z$, explaining the difference to the soft limit.

As illustrated in Fig.~\ref{fig:diagram}, the evolution in the medium also results in the production of stimulated \textit{bremsstrahlung} radiation, in parallel to the virtuality cascade. This mechanism has two significant phenomenological consequences: the enhancement of the splitting function at large angles and radiative energy loss to the matter. We now describe how these effects can be incorporated into the evolution of the track functions.

The multiple soft momentum gluons exchanged with the medium result in a characteristic interference pattern, leading to an enhanced branching rate, first computed by Landau, Pomeranchuk, and Migdal in QED~\cite{Landau:1953um,Migdal:1956tc,Klein:1993vq}. This regime, captured in the BDMPS-ASW approximation, is valid for intermediate energies of the radiated partons, $\omega_{\rm BH}<z(1-z)p_t<\omega_c$, where $\omega_{\rm BH}$ denotes the Bethe-Heitler frequency, below which the radiative spectrum is dominated by incoherent contributions~\cite{Bethe:1934za}, and $\omega_c = \frac{1}{2}\hat q L^2$ denotes the characteristic frequency above which the gluon resolves the medium coherently. A closed-form for the radiation rate is not known in QCD in this region of phase space,  and we consider an asymptotic limit of exactly collinear branchings followed by in-medium diffusion:\footnote{See e.g.~\cite{Barata:2021wuf} for further discussion on this approximation.}
  \begin{align}\label{eq:iop}
   \frac{dI}{ dz d\mu \, z^2(1-z)^2\mu} = \int_0^L dt \,  \frac{\cP(\mu z(1-z) ,L-t) }{2\pi}\frac{dI}{dz dt}\, ,
\end{align}
where $dI/dzdt$ denotes the medium induced radiation rate and $\cP$ is the so-called momentum broadening distribution. The transverse momentum is only acquired due to the evolution between the time the outgoing states go on-shell and the end of the medium; the in-medium rate determines the energy distribution. In the BDMPS-Z/ASW approximation the broadening distribution is purely diffusive~\cite{Blaizot:2013vha, Barata:2020rdn}, $\cP(k,t) = 4\pi/(\hat q_{\rm eff} t) \exp(-k^2/(\hat q_{\rm eff}t))$; the radiative rate reads~\cite{Blaizot:2013hx,Blaizot:2013vha}
\begin{align}\label{eq:spec:energy}
    \frac{dI}{ dz dt}
    &= \frac{\alpha_s N_c}{2\pi}  \frac{f^{\frac{5}{2}}(z)}{(z(1-z))^{\frac{3}{2}}}\sqrt{\frac{\hat q }{p_t}}\, .
\end{align}
Eq.~\eqref{eq:spec:energy} can be regularized by a plus distribution while shifting the singularities to $z=1$, and using the integral identity~\cite{Blaizot:2013vha} 
\begin{align}
 \int_\varepsilon^{1-\varepsilon} dz\,    \frac{dI}{dz dt} = \int_0^{1-\varepsilon} dz\,  2z \, \frac{d I}{ dz dt}\, .
\end{align}
Combining all these elements, Eq.~\eqref{eq:iop} becomes 
\begin{align}
   &\frac{dI}{dz d\mu} = \frac{\alpha_s N_c}{\pi} \sqrt{\frac{1}{\hat q\,  p_t}} \mu   \nn 
   &\times \frac{(2z) \, z^2(1-z)^2 f^{\frac{3}{2}}(z)}{(z(1-z))^{\frac{3}{2}}} \Gamma_0\left(\frac{z^2(1-z)^2\mu^2}{f(z) \hat q L}\right)\, ,
\end{align}
where $\Gamma_0$ denotes the incomplete Gamma function.

\begin{figure}[h!]
    \centering
    \includegraphics[width=.45\textwidth]{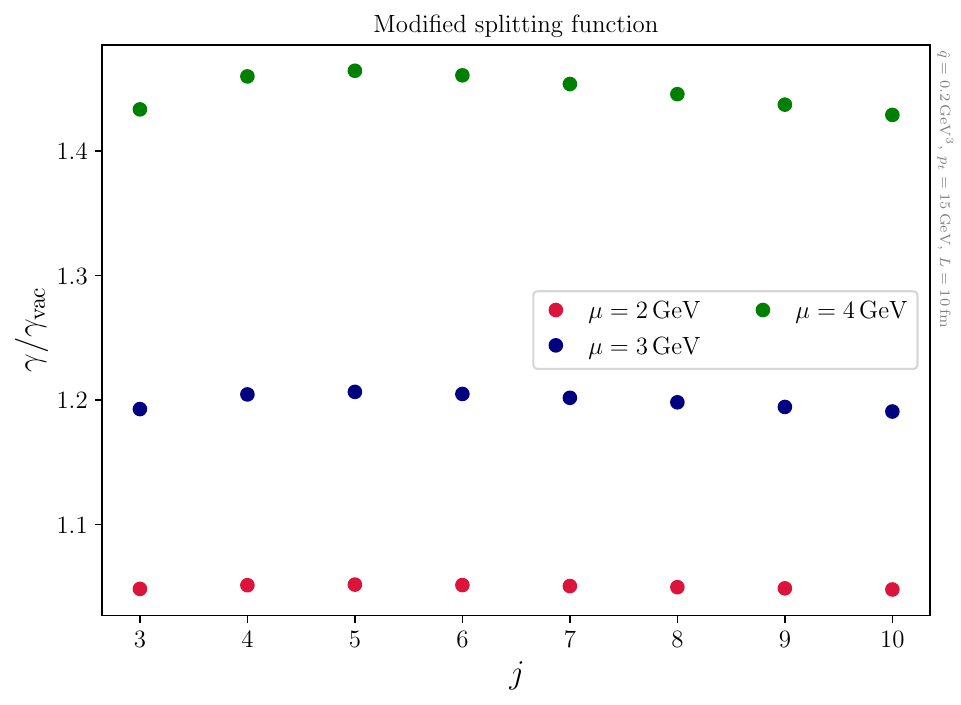}
    \caption{Anomalous dimension evolution including medium induced radiation.}
    \label{fig:gamma_kernel}
\end{figure}

We include the medium-induced kernel in the splitting functions entering the RG evolution in Eq.~\eqref{eq:track_general}. Such a choice has been applied in different jet quenching phenomenological studies, see e.g~\cite{Ke:2023ixa,Deng:2009ncl,Majumder:2013re, Armesto:2009fj,Chien:2015vja}. 
Note that the in-medium effects impact both the anomalous dimensions and matrix elements of the operators, i.e., the boundary conditions for the evolution. We start the evolution from the hard scale, where the in-medium effects can be treated perturbatively. Thus, in an effective field theory spirit, the hard scale is assumed to be parametrically much larger than medium-induced scales, and the dominant effect comes from the evolution, i.e. from in-medium modification of the anomalous dimensions. A complete effective field theory formulation of in-medium effects is still missing; although such an effort would be a worthwhile endeavor, it goes past the scope of this study, and we leave it for future work. Thus, the anomalous dimensions in Eq.~(\ref{eq:AD}) become
\begin{align}\label{eq:gamma_med_kernel}
   &\gamma(j,\mu)=  -\int_0^1 \Bigg\{P(z) +   N_c \,  \mu^2 \sqrt{\frac{1}{\hat q\,  p_t}}  \nn 
   &\times  2z \sqrt{z(1-z)}\,f^{\frac{3}{2}}(z)  \Gamma_0\left(\frac{z^2(1-z)^2\mu^2}{f(z)\hat q L}\right) \Bigg\}  z^{j-1}\, ,
\end{align}
up to a $\delta(z-1)$ term, whose coefficient can be obtained by imposing the sum rule $\gamma(2)=0$. In Fig.~\ref{fig:gamma_kernel} we evaluate Eq.~\eqref{eq:gamma_med_kernel}. To ensure that the collinear cascade approximation is valid, we require $p_t \ll \hat q L^2$~\cite{Blaizot:2013hx}. Therefore, we take a long medium with $L=10$ fm and a small jet energy with $p_t=15$ GeV. The results show a mild dependence on $j$, while the evolution in $\mu$ is fast, leading to a substantial enhancement that competes with the phase space constraint effect in Fig.~\ref{fig:gamma_PS}. The magnitude of the increase is related to the asymptotic approximation used; although other calculations of the in-medium rate at finite $z$ exist~\cite{Isaksen:2020npj,Isaksen:2023nlr,Dominguez:2019ges,Apolinario:2014csa,Blaizot:2012fh}, it is unclear how to incorporate them in the RG evolution at present.\footnote{This is possible for thin or dilute matter, see~\cite{Sievert:2018imd}.} Due to the high sensitivity to the modeling, we will not discuss these contributions further; nonetheless, we expect this modification to yield a small effect due to the absence of a collinear pole in the medium.

So far, we have considered the contribution of \textit{bremsstrahlung} radiation, which hadronizes into charged hadrons. However, a substantial part of the softer induced radiation feeds down to the medium scale, and it should be viewed as part of the QGP while quenching the jet energy. We describe these \textit{lost} emissions in terms of a distribution function $D(\varepsilon)$, with $\varepsilon \ll p_t$ representing the amount of quenched energy, see Fig.~\ref{fig:diagram}. This distribution can be computed perturbatively under certain limits~\cite{Baier:1994bd,Mehtar-Tani:2017ypq,Mehtar-Tani:2016aco}, but it is expected to have a significant non-perturbative contribution. Therefore, we remain agnostic to its functional form and only impose overall energy conservation, i.e., $\int_0^\infty d\varepsilon D(\varepsilon)=1$.

To implement the effect of energy loss in the TFs' RG, we first consider the simpler case of DGLAP evolution; at $\mathcal{O}(\alpha_s^2)$ the real contribution to the gluon distribution $G$ reads~\cite{Altarelli:1977zs}
\begin{align}\label{eq:iioii}
  &G(x,\mu) \Big\vert_{\rm real}^{\mathcal{O}(\alpha_s^2)}= \frac{\alpha_s^2}{(2\pi)^2 }\int_{z_1,z_2,x_1} \, \hat P(z_1) \hat P(z_2) \nn 
  &\times\int_{\mu_0}^{\mu} d \ln\mu'\int_{\mu_0}^{\mu'} d \ln\mu'' G(x_1,\mu'') \delta(x - x_1 z_1 (1-z_2))  \, ,
\end{align}
  where $\hat P(z)= \hat P(1-z)$ is the unregularized splitting function. We choose $z_2$ to represent the fraction of energy lost to the matter. The RHS of Eq.~\eqref{eq:iioii} can be manipulated the following way (dropping the $\mu$ dependence):
  \begin{align}
  &\frac{\alpha_s^2}{(2\pi)^2 }\int_x^1 dz_1 \int_0^{1-\frac{x}{z_1}} dz_2 \, \frac{\hat P(z_1) \hat P(z_2)}{z_1(1-z_2)} G\left(\frac{x}{z_1(1-z_2)}\right)\nn
  &\approx \frac{\alpha_s}{2\pi }\int_x^1 \frac{dz_1}{z_1} \int_0^{(z_1-x)p_t} d\varepsilon\,  \hat P\left(z_1+\frac{\varepsilon}{p_t}\right) D\left(\varepsilon\right) G\left(\frac{x}{z_1}\right)\, ,
  \end{align}
  where we replaced the second splitting function by the energy loss distribution $D(\varepsilon)$, modeling energy transferred to the matter, see Fig.~\ref{fig:diagram}. Assuming $ 1>x~\sim z_1\gg \frac{\varepsilon}{p_t}$, the upper cut-off of the $\varepsilon$ integration should be extended to infinity. Treating the energy loss as an additional convolution within the standard DGLAP cascade  and including the virtual corrections, we find   
  \begin{align}
  \mu \frac{d G(x,\mu)}{d \mu}&\approx \frac{\alpha_s}{2\pi}\int_{x_1,z} \int_0^\infty d\varepsilon \, D(\varepsilon)  P\left(z + \frac{\varepsilon}{p_t}\right) \nn 
  &\times G(x_1) \delta(x-x_1 z ) \, ,
\end{align}
  which agrees with the Sudakov factor found in~\cite{Mehtar-Tani:2016aco}. A similar procedure for the TFs, with the initial truncation in Eq.~\eqref{eq:iioii} at $\mathcal{O}(\alpha_s^3)$, with $ \delta(x-x_1 z (1-z_2)) \to \delta(x-x_1 z_1 (1-z_2) - x_2(1-z_1)(1-z_3))$, where $z_2,z_3$ are the energy loss fractions from uncorrelated emissions, leads to
  \begin{align}
    &\mu\frac{d T}{d \mu}  =  \frac{\alpha_s}{2\pi}\int_{\varepsilon_1,\varepsilon_2} D(\varepsilon_1) D(\varepsilon_2)\int_0^1 dz P\left(z + \frac{\varepsilon_1}{p_t}+ \frac{\varepsilon_2}{p_t}\right)   \nn 
    &\int_{x_1,x_2} T(x_1,\mu) T(x_2,\mu)  \, \delta(x-zx_1-(1-z)x_2)\, . 
  \end{align}
The upper limit of the $\varepsilon_{1,2}$ integrals is set to infinity. The evolution for the moments at LO is unchanged with respect to Eq.~\eqref{eq:track_general} if the anomalous dimensions include the double convolution with the energy loss probabilities.
  
\begin{figure}[h!]
    \centering
    \includegraphics[width=.45\textwidth]{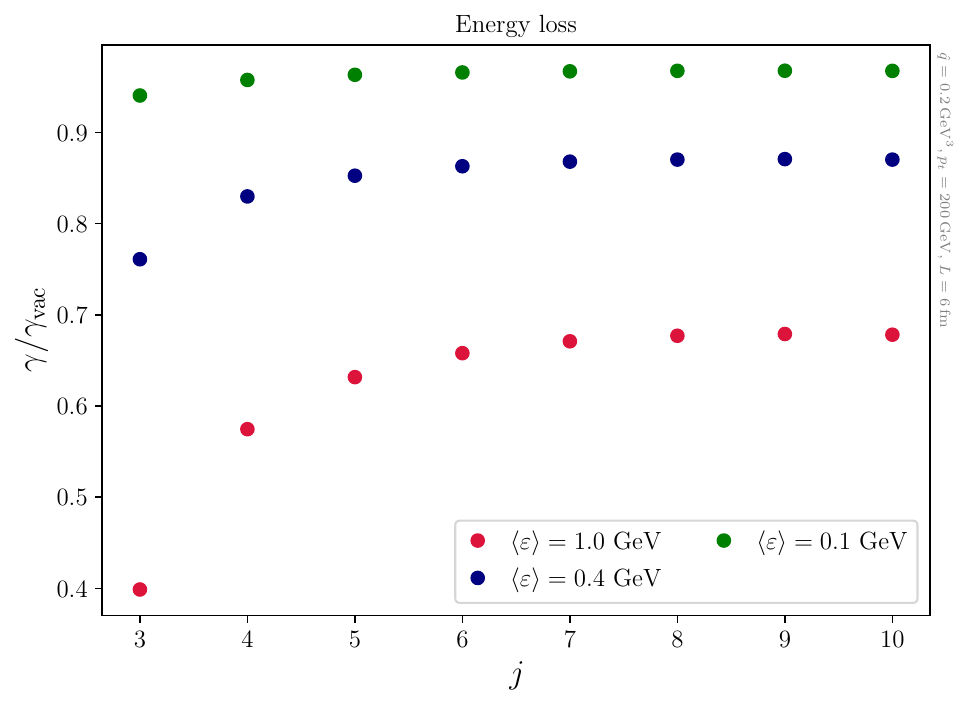}
    \caption{Anomalous dimension evolution including average energy loss effects. We use the same parameters as in Fig.~\ref{fig:gamma_PS}.}
    \label{fig:gamma_eloss}
\end{figure}

Without specifying the form for $D(\varepsilon)$, but assuming that the energy loss distribution has a small dispersion,  we simplify the evolution equation replacing $\varepsilon$ by its mean value $\langle \varepsilon \rangle\sim \hat q L^2$. We note that this approximation might be insufficient for phenomenological applications, see e.g.~\cite{Baier:2001yt}. With this caveat, the anomalous dimension reduces to 
\begin{align}
   \gamma(j)    &= -\int_{\frac{2\langle \varepsilon\rangle }{p_t}}^1 dz \, P(z)  \left(z -\frac{2 \langle\varepsilon \rangle}{p_t}   \right)^{j-1}\, ,
\end{align}
up to a $\delta(1-z)$ term. We can no longer fix its coefficient via the sum rule $\gamma(2)=0$ since we assumed $1>x\gg \frac{\varepsilon}{p_t}$. Nonetheless, we impose the sum rule in Fig.~\ref{fig:gamma_eloss} to  directly compare energy-loss driven modifications to Figs.~\ref{fig:gamma_PS} and~\ref{fig:gamma_kernel}. Compared to the naive estimation from the BDMPS-Z/ASW approximation, $\langle \varepsilon \rangle_{\rm BDMPS-Z/ASW}\sim \mathcal{O}(20)$ GeV, we observe significant effects already for $\langle \varepsilon \rangle$ one order of magnitude lower (red markers). This is likely due to the mean value approximation, which typically overestimates the quenching effects~\cite{Baier:2001yt}. Consequently, we restrict ourselves to small values for $\langle \varepsilon \rangle/p_t$, such that non-physical results are avoided. 

\textit{Application to EECs:} Having considered the modifications to the TFs' RG evolution, we study their impact on the jet EEC. The LO EEC on tracks is defined as~\cite{Chen:2020vvp,Li:2021zcf} 
\begin{align}\label{eq:EEC_main} 
 \frac{d\Sigma^{(n)}_{\rm trk}}{d\chi} &\equiv    \int_{z,x_1,x_2} x_1^n T(x_1)  x_2^n T(x_2) z^n(1-z)^n \frac{d\sigma}{\sigma dz d\chi}\nn 
&=\int_z  T^{[n]}(\chi p_t)   T^{[n]}(\chi p_t) z^n(1-z)^n \frac{d\sigma}{\sigma dz d\chi} \, ,
\end{align} 
where we only include terms for $\chi>0$. In YM theory, the lowest order non-trivial track function appears at $n=2$; see Eq.~\eqref{eq:T1_moment_vac_gamma}, which we consider in what follows. We work at fixed coupling, such that we are only sensitive to the evolution of the track functions, and  for the initial condition, we take the simplistic form $T(x,\mu_0)=252x^2(1-x)^6$ at $\mu_0=p_t$~\cite{Chang:2013iba}. Equation~\eqref{eq:EEC_main} is computed using the in-medium cross-section obtained in the limit of highly energetic final states, see~\cite{Dominguez:2019ges,Isaksen:2020npj,Isaksen:2023nlr} for details. Writing $d\sigma =  d\sigma_{\rm vac} \left(1+F_{\rm med}\right)$, we have
\begin{align}\label{eq:Fij}
	F_{\rm med}&= \frac{2}{t_f} \Bigg\{\int_0^L d t\,  \int_t^L \frac{d t'}{t_f} \,  \left[\cos\left(\frac{t'-t}{t_f}\right) \mathcal{C}_3(t',t) \mathcal{C}_4(L,t')\right] \nn 
 &- \sin\left(\frac{L-t}{t_f}\right) \mathcal{C}_3(L,t)\Bigg\}\, .
\end{align}
The correlators $\mathcal{C}_{3,4}$ in the BDMPS-Z/ASW approximation take the form $\mathcal{C}_{3}(t',t)\approx e^{-\frac{1}{12} \hat q \chi^2 \ (t'-t)^3}$ and $\mathcal{C}_4(L,t')\approx e^{-\frac{1}{4}\hat q \chi^2 (L-t') (t'-t)^2}$, where we neglected the $z$ dependence in the phases.

\begin{figure}[]
    \centering
    \includegraphics[width=.45\textwidth]{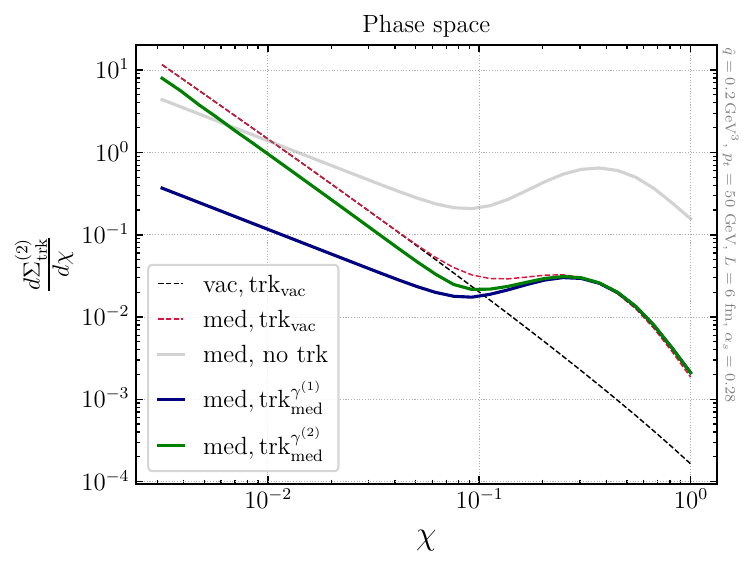}
    \includegraphics[width=.45\textwidth]{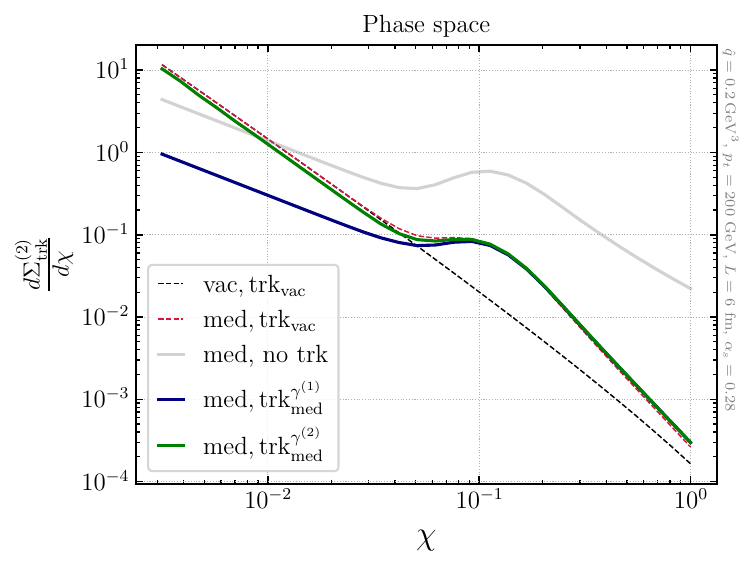}
    \caption{Result for the EEC with $n=2$, including phase space effects. Top: $p_t=50$ GeV; Bottom: $p_t=200$ GeV. }
    \label{fig:EEC_PS}
\end{figure}

\begin{figure}[]
    \centering
    \includegraphics[width=.45\textwidth]{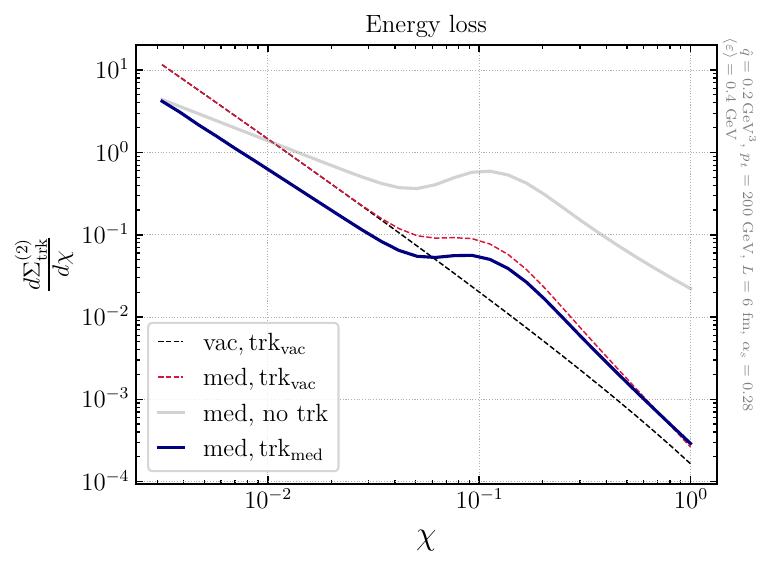}
    \includegraphics[width=.45\textwidth]{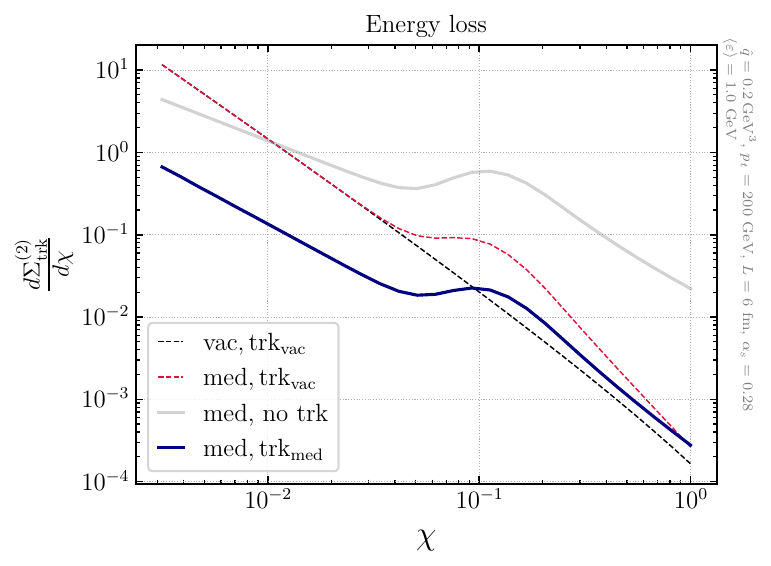}
    \caption{Result for the EEC with $n=2$, including mean energy loss. Top: $\langle \varepsilon \rangle=0.4$ GeV; Bottom: $\langle \varepsilon \rangle=1$ GeV.  }
    \label{fig:EEC_eLoss}
\end{figure}

The results for $\Sigma^{(2)}_{\rm trk}$ are shown in Figs.~\ref{fig:EEC_PS} and \ref{fig:EEC_eLoss}. For the case of reduced phase space, including the medium-modified track functions leads to small changes compared to vacuum TFs. The major differences arise at smaller values of $p_t$, in the region where the medium leads to an enhancement of the EEC. Importantly, including the in-medium tracks does not alter the shape of the distribution, which allows us to qualitatively visualize the color decoherence transition. The distinction between the phase space models is clear for small $\chi$, where out-of-the-medium fragmentation becomes crucial. Mean energy loss shifts the distribution below the vacuum case, as is typical of energy loss effects in jet observables. Again, the overall shape of the distribution is only barely modified, even for the largest $\langle \varepsilon \rangle$. 

\textit{Conclusions:} We conducted the first study of the track functions RG evolution in a dense QGP environment. Comprehending such effects is vital in heavy ions, where soft, uncorrelated sources contaminate the jet EEC. We showed that, at a qualitative level, it is adequate to use the vacuum TFs as the QGP medium does not drastically impact the shape of the EEC distribution through the modified TFs. 

Our analysis is phenomenologically driven and faces several shortcomings, such as the need to describe the phase space for vacuum-like emissions at leading logarithmic accuracy and the lack of suitable treatment of the in-medium matrix elements. From a phenomenological point of view, it would be worthwhile to extract the in-medium TFs from available data or MC samples. We hope to address these aspects in the future.

\textit{Acknowledgments:} JB and RS are supported by the United States Department of Energy under Grant Contract DESC0012704. We are grateful to Swagato Mukherjee, Guilherme Milhano, Yacine Mehtar-Tani, Ian Moult, Wenqing Fan, Andrey Sadofyev, Xin-Nian Wang, and Mateusz Ploskon for discussions. We particularly thank Paul Caucal, Pier Monni, and Alba Soto-Ontoso for their collaborative efforts on closely related work and helpful discussions.

\bibliographystyle{apsrev4-1}
\bibliography{references.bib}

\end{document}